\documentclass[oneclomun]{aa}
\usepackage{graphicx}
\usepackage[authoryear]{natbib}
\bibpunct{(}{)}{;}{a}{}{,}

\def\ltsima{$\; \buildrel < \over \sim \;$}
\def\simlt{\lower.5ex\hbox{\ltsima}}
\def\gtsima{$\; \buildrel > \over \sim \;$}
\def\simgt{\lower.5ex\hbox{\gtsima}}
\def\simlt{\lower.5ex\hbox{\ltsima}}

\def\kms{\ifmmode {\rm \ km \ s^{-1}}  
\else
$\rm km \ s^{-1}$\fi}

\def\h2{\mbox{\ion{H}{ii}}}

\newcommand{\lessim}{\mathrel{\hbox{\rlap{\hbox{\lower4pt\hbox{$\sim$}}}\hbox{$<$}}}}

\begin{document}
\addtolength{\voffset}{1cm}


\title{Constraints on the star formation history of the Large Magellanic Cloud}

\author{S. C. Javiel \and B. X. Santiago \and L. O. Kerber }

\offprints{S. C. Javiel, \email{basagran@if.ufrgs.br}} 

\institute{Universidade Federal do Rio Grande do Sul, IF, 
CP\,15051, Porto Alegre 91501--970, RS, Brazil}

\date{Received / Accepted 1 October 2004}

\titlerunning{Star Formation History of the LMC}
\authorrunning{S. C. Javiel}
\abstract{We present the analysis of deep colour-magnitude diagrams (CMDs) of 
6 stellar fields in the LMC. The data were obtained using HST/WFPC2 in the 
F814W ($\sim$ I)  and  F555W ($\sim$ V) filters, reaching $\mathrm{V_{555}
\sim 26.5}$. We discuss and apply a method of correcting CMDs for
photometric incompleteness. A method to generate artificial
CMDs based on a model star formation history is also developed. 
This method incorporates photometric error effects, unresolved binaries,
reddening and allows use of different forms of the initial mass function
and of the SFH itself. We use the Partial Models Method, as presented
by Gallart and others, for CMD
modelling, and include control experiments to  prove its validity in a
search for constraints on the Large Magellanic Cloud star formation history
in different regions. Reliable star formation histories for
each field are recovered by this method. In all fields, a gap in 
star formation with $\tau \sim$ 700 Myrs is observed. Field-to-field
variations have also been observed. The two fields 
near the LMC bar present some significant
star forming events, having formed both young ($\tau$ \ltsima $1~Gyr$) 
and old ($\tau$ \gtsima $10~Gyr$) stars, with a clear gap from
$3-6$ Gyrs. Two other fields display quite similar SFHs, with increased star
formation having taken place at $\tau \simeq 2-3 Gyrs$ and $6$ \ltsima $\tau$
\ltsima $10 Gyrs$. The remaining two fields present star formation histories 
closer to uniform, with no clear event of enhanced star formation.
\keywords{galaxies: galaxies: formation, galaxies: evolution; stars: statistics}}

\maketitle

\section{Introduction}
\label{intro}
The study of the star formation history (SFH) in near galaxies can
contribute to our understanding of the evolution of these objects, the 
mechanisms involved in the star formation bursts, their associated 
time scales, and the initial stellar mass function (IMF). 
Star formation time scales and IMFs are particularly important, being
used as input for population synthesis models which try to interpret 
integrated light from distant galaxies to obtain information about
their component stars. Moreover, by matching SFH to dynamical models, 
it is possible to investigate the
influence of gravitational interactions on star formation.
 
The Large Magellanic Cloud (LMC) is a satellite of the Galaxy, being
at the same time quite distinct from it and yet near enough to it 
to allow a study of its structure, internal kinematics and 
stellar populations (Westerlund 1990). By its
proximity and nature, the LMC provides an excellent laboratory for
the study of the processes involved in star formation,
such as interactions with neighboring galaxies 
(especially the Galaxy and the Small Magellanic Cloud (SMC)), the influence of bar dynamics 
and chemical enrichment (Gardiner et al. 1998).

The main advantage of LMC stellar population studies is the ability to 
build detailed colour-magnitude diagrams (CMDs) in various regions within it.
Furthermore, the LMC supplies a unique chance to compare the inferred 
SFH through the study of field stars with the SFH found from studies of its
cluster system. The reason is that the LMC presents a rich cluster
population of largely varying ages. Cluster CMDs have the advantage
of being composed of a single population, making their analysis
much simpler. Studies
using LMC clusters show that the majority of them are relatively young,
with ages $\tau$ \simlt 4 Gyr, a small number has $\tau$ 
\simgt 10 Gyr and only one is of intermediate age ($\tau \sim $ 8
Gyr) (Mateo, Hodge \& Schommer 1986). However, only a
small fraction of LMC stars belong to clusters, rendering the field
CMD analyses of more general interest and more representative
of the global SFH.

Ground-based CMD studies have favoured a predominance of young
or intermediate-age populations in the LMC. 
Bertelli et al. (1992), for example, 
have analyzed CMDs in a few LMC outer regions and modelled them with a 
single-burst SFH with $\tau \sim$ 2 -- 4 Gyr, with a
factor of $\sim 10$ increase relative to the average quiescent star
formation, the exact burst age being dependent on the details of
stellar evolutionary theories available then. The same group also
found some evidence for field-to-field variations in SFH (Vallenari et al. 1996).

Earlier ground-based work did not reach deep enough magnitudes 
to directly access the main sequence turn-offs of stellar
populations with ages $\tau$ \simgt 6 Gyr. 
This became possible only with the use of the Hubble Space Telescope
(HST), whose images were free of atmospheric turbulence and,
therefore, much sharper and deeper. Gallagher et
al. (1996), based on a single HST Wide Field and Planetary Camera 2
(WFPC2) field, suggested a roughly constant 
star formation rate (SFR) in the past few Gyr, except for a small
period of enhanced star formation about 2 Gyr ago. 
Holtzman et al. (1997) analyzed the
luminosity function of the same HST field, located some 4$\degr$
from LMC bar, and favoured a constant SFR for 10 Gyr in the LMC, followed
by an increase factor of 3 in the last 2 Gyr, yielding roughly the
same number of stars younger and older than 4 Gyr. They also
identified an intermediate-age turn-off in their field CMD. Elson, Gilmore
\& Santiago (1997) studied an HST field closer to the LMC bar and
identified two main populations, with ages $\tau \, \sim$ 1 -- 2 Gyr
and $\tau \, \sim$ 2 -- 4 Gyr, the exact ages depending on
their metallicity. Geha et al. (1998) have studied 3 HST fields in the outer
region of the LMC and also suggested a model with roughly equal
numbers of stars younger and older than $\tau =$ 4 Gyr. Studying fields
inside the LMC bar, Olsen (1999) suggested that the star formation has 
been more continuous, possibly extending back for a longer period and taking 
place at a roughly constant SFR as compared to the outer fields. This 
differs from the conclusions of
Elson, Gilmore \& Santiago (1997). Holtzman et al. (1999) have 
not found obvious evidence for bursts of star formation and concluded
that the field SFH differs from that based on the LMC cluster age
distribution.

In this paper, we describe the CMD modelling of 6 deep HST fields
located in different directions along the LMC, all of them 
within $6 ^\circ$ of the 
LMC centre. Our main goal is to recover
information about the SFH of the LMC field population, the 
age-metallicity relation $Z(\tau)$ and IMF. Our samples have been
combined with those of a previous paper (Castro et al. 2001) and make
up an extended, deep and homogeneous photometric dataset.
The paper is outlined as follows: in Sect. 2 we
describe the HST/WFPC2 data, the stellar samples derived from them, 
and the photometry, including methods of accounting for sample
incompleteness. The combined sample is presented in Sect. 3; in Sect. 4
we present the CMD modelling algorithm, which is
essentially the Partial Models Method (PMM) by Gallart et al. (1999), 
Aparicio et al. (2001) and Carrera et al. (2002), and discuss the statistical
tools used in our model vs. data comparisons; finally, in Sect. 5 we 
present the results, which are discussed in Sect. 6.

\section{Observations and data reduction}

We use data obtained with the {\it Wide Field and Planetary Camera 2}
(WFPC2) on board HST for 6 fields near the rich LMC clusters 
NGC1805, NGC1818, NGC1831, NGC1868, NGC2209 and Hodge 11. 
These data are part  of the GO7307 project entitled ``Formation and
Evolution of Rich Star Clusters in the LMC'' (Beaulieu et al. 1999;
Kerber et al. 2002; de Grijs et al. 2002). For each field, a set of
at least 2 exposures were obtained using each of the F814W (I) and F555W (V)
filters.  The main properties of the fields 
(hereafter identified by their nearby rich LMC cluster) are given in
Table \ref{tab1}: columns 2 and 3 list their equatorial coordinates (J2000),
followed by their angular distance from the LMC optical centre, total
I and V exposure times and the number of individual exposures taken
with each filter.

All exposures were put through the standard HST pipeline procedure
that corrects them for several instrumental effects, such as bias and
dark currents and flat-fielding (Holtzman et al. 1995a). The exposures
taken in each field/filter configuration were then combined using the
IRAF\footnote{IRAF is the Image Reduction and Analysis Facility, a general
purpose software system for the reduction and analysis of astronomical data. IRAF is written and supported by the IRAF programming group at the National Optical Astronomy Observatories (NOAO) in Tucson, Arizona. NOAO is operated by the  Association of Universities for Research in Astronomy (AURA), Inc. under cooperative agreement with the National Science Foundation} task CRREJ, in order to increase the signal to noise ratio and remove
cosmic rays. The final combined image at each configuration was then
used for sample selection and photometry.

\begin{table}
\caption[]{Basic characteristics of the our HST/WFPC2 fields. 
The columns show the field name, equatorial coordinates
(J2000), angular distance with respect to the LMC centre, total exposure 
times and the number of exposures in I and V. Note that the identification 
of each field is related to the rich LMC cluster parallel to which it was 
observed.}
\label{tab1}
\small
\renewcommand{\tabcolsep}{0.9mm}
\begin{center}
\begin{tabular}{c c c c c c c c}
\hline
Field & $\alpha$ (h m s) & $\delta$ ($\deg~\arcmin~\arcsec$) & Ang.  & $t_{I}$ & $t_{V}$ & $N_{I}$ & $N_{V}$  \\

$~~~~$ & $(2000)$ & $(2000)$ & Dist. ($\degr$) & $(s)$ & $(s)$ & $~~~~$ & $~~~~$  \\
\hline 
NGC1805 &  5 01 42 & -65 59 58 & 4.28 & 2200 & 2200 & 2 & 2  \\
NGC1818 &  5 05 00 & -66 25 13 & 3.75 & 4800 & 7200 & 4 & 6 \\
NGC1831 &  5 05 21 & -64 49 50 & 1.51  & 2200 & 2200 & 2 & 2  \\
NGC1868 &  5 13 48 & -63 52 10 & 5.96  & 2200 & 2200 & 2 & 2  \\
NGC2209 &  6 07 39 & -73 46 07 & 5.28 & 4800 & 7200 & 4 & 6 \\
Hodge 11 & 6 15 07 & -69 49 08 & 4.44  & 4800 & 7200 & 4 & 6  \\
\hline
\end{tabular}
\end{center}
\hspace{3cm}
\end{table}

\subsection{Sample selection}

The IRAF DAOPHOT package was used to detect and classify sources
automatically in each combined image, i. e. the sample selection was
carried out independently in each photometric band. Star candidates
were detected using DAOFIND, with a peak intensity threshold for
detection set to $5\,\sigma$, where $\sigma$ corresponds to the rms 
fluctuation in the sky counts, determined individually for each
combined image.

\begin{figure}[!h] 
\resizebox{\hsize}{!}{\includegraphics{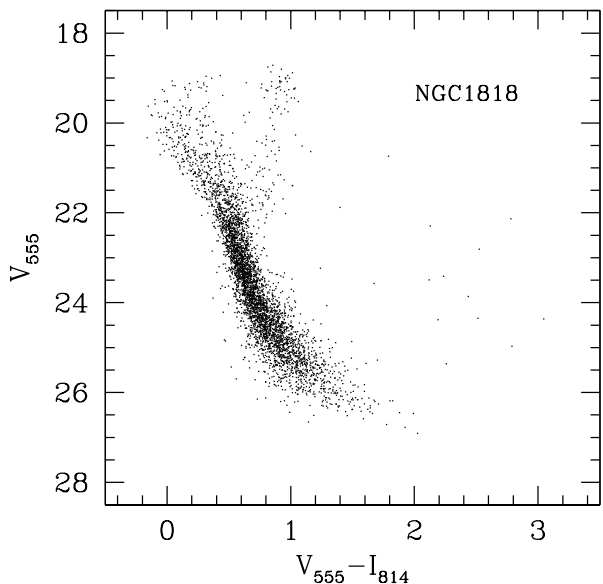}}
\caption[]{Observed colour-magnitude diagram of the field near NGC1818.} 
\label{cmd1818}
\end{figure}

A preliminary aperture ($radius = 2 \arcsec$) photometry was carried
out by running the task PHOT on all objects output by
DAOFIND. Star/galaxy separation required modelling point sources by
fitting the profiles of bright, isolated and non-saturated stars to a
Moffat function ($\beta = 1.5$). We used the IRAF task PSF for that
purpose. This PSF model was then fitted to all objects found by
DAOFIND with the ALLSTAR task.

The output parameters given by ALLSTAR ($\chi^2$, sharpness, magnitude
uncertainties $\mathrm{\delta I}$ and $\mathrm{\delta V}$) are
useful to separate a purely stellar sample from the remaining detected sources,
like distant galaxies. Stellar sources present small $\mathrm{\delta
I}$ and $\mathrm{\delta V}$ values. Thus we threw away objects 
with $\mathrm{\delta I}$ or $\mathrm{\delta V}$ greater than a
constant value, typically $\sim$ 0.15 mag. These cut off values
were chosen separately for each field/filter, based on visual
inspection of the more conspicuous stellar and non-stellar cases.
This stellar/non-stellar classification was checked by visual
inspection of the stellar sample selected and iterated when necessary,
until a ``clean'' stellar sample resulted. The final stellar sample
in each field, and which was used in the subsequent CMD analysis and modelling,
was the merging result of stars selected in both bands, i. e., with
small values of $\mathrm{\delta I}$ and $\mathrm{\delta V}$ measured
by ALLSTAR.

Figure \ref{cmd1818} shows the final observed CMD of the field near the
cluster NGC1818. One can see a main sequence (MS) that stretches down to 
$\mathrm{V_{555}\sim V=26.5\,}$. The saturation limit,
also found using the ALLSTAR $\mathrm{\delta I}$ and 
$\mathrm{\delta V}$ parameters (saturated stars display large
magnitude uncertainties from PSF fitting), was found to be about 
$\mathrm{V_{555}=18.5}$ for this field. The red giant branch (RGB) is
visible as well, along with a turn-off around $\mathrm{V_{555} \simeq 22.5\,}$
This latter corresponds to an old population ($\tau$ \simgt 10 Gyr, as
shown in Castro et al. 2001). The red giant clump (RC), found in
intermediate-age populations, is also visible  at $\mathrm{V_{555}\sim 19}$
and $\mathrm{V_{555}-I_{814}\sim 0.9\,}$. The faint red stars at
the right of the CMD are likely low-luminosity M dwarfs in the
Galaxy. The other 5 deep LMC fields in our sample display the same
general features described above.

The data suffer from an important effect: sample incompleteness. Our 
CMD modelling algorithm (presented in \S 4) therefore needs to 
incorporate such an effect in order to place models and data on equal
footing. Quantifying photometric uncertainties and 
applying them to the model CMDs is extremely important as well, as
they are responsible for most of the observed CMD spread. These data
corrections are the subject of the next subsections.

\subsection{Photometric uncertainties}

There are two approaches to quantify photometric uncertainties: an
empirical approach, through the comparison of independent magnitude 
measurements of the
same object, or a semi-empirical approach, in which the photometric
uncertainties follow from the measured signal and an adopted model 
for the associated noise, such as that used by the IRAF task PHOT. 
In the empirical approach, one needs two or more 
images of identical characteristics, such as total exposure time, 
detector noise, etc. We unfortunately do not have two identical combined
images for each field and filter.

We tried to use the independent magnitude measurements 
taken from individual exposures of each filter/field configuration in order to
empirically estimate the photometric uncertainties in the combined
images.  However, the photometric error
measured in a single exposure tends to be greater than that of the
combined image, due to the smaller signal to noise ratio. 
Assuming Poissonic statistics of photon counts, overlooking
detector noise and assuming $N$ individual exposures of the same exposure time,
the scaled magnitude uncertainty of the combined image should be:

\begin{equation}
\delta m \propto \left( \frac{1}{\sqrt{N}}\right)
\label{eq_scaling}
\end{equation}

However, we failed to confirm the scaling in magnitude uncertainties
given above, which suggests that
the assumptions that lead to eq. \ref{eq_scaling} are over-simplistic.
Therefore, we decided to rely on the model uncertainty
from the IRAF PHOT task. These uncertainties take detector (readout) noise and
background (sky) noise into account as well. The spread of stars
in the CMD depends on the photometric uncertainties of both
bands. This means that the expected 
spread in colour $\mathrm{\sigma_{colour}}$ is given by
\begin{equation}
\mathrm{\sigma_{colour}^2 =\sigma_{I}^2+ \sigma_{V}^2 }
\label{err_eq1}
\end{equation}

\subsection{Sample completeness}

In any image, bright stars, which are not the majority, are
easily detected. The faint ones are often not detected. 
Photometric completeness is defined as the percentage
of the total number of stars that were successfully 
included in the sample by the detection and
selection process. The knowledge of the completeness function is essential when
the objective is to compare an observed CMD to model ones, with the goal
of reconstructing an SFH. We thus carried out experiments to estimate
the completeness function for each of our HST/WFPC2 fields. These
experiments involved adding the same artificial stars,
with a given magnitude and color, to the combined V and I images,
and submitting these images, with real plus artificial stars, to the
same detection and selection process as the real data. Thus, the 
extra stars were subjected to automatic detection with DAOFIND,
to aperture photometry using PHOT, to PSF fitting and 
star/no-star separation using
ALLSTAR and, finally, to the final matching of stellar lists
in the I and V bands, as described in Sect. 2.1. Most
previous works estimate completeness as a single variable function,
such as C(V) or C(I), often overlooking 
variations in completeness along the color axis in a CMD. 
Others have computed single variable functions independently and have
then taken their product in order to quantify completeness
at any given CMD position (Mateo 1988). However, since the position within the
field (and therefore the proximity to a bright star, for instance) is
the same in both filters, the detection of a given star in the
combined V image is not an event independent of detection of the
same star in the combined I image. Thus, this latter approach
invariably leads to an underestimate of completeness, the amplitude
of the bias depending on issues such as field crowding, exposure
times, etc. Our approach, by assigning a
magnitude and colour to each artificial star and keeping the same CCD position
for both filters, more effectively reproduces the true bi-variate
completeness function. This resulting 
completeness function is then expressed as:

\begin{equation}
\mathrm{C(V-I,V)=\frac{recovered\: artificial \: stars\: in\: both\: V\: and\: I}{input \:artificial \:stars }}
\label{comp_eq1}
\end{equation}
   
\begin{figure}[!h] 
\resizebox{\hsize}{!}{\includegraphics{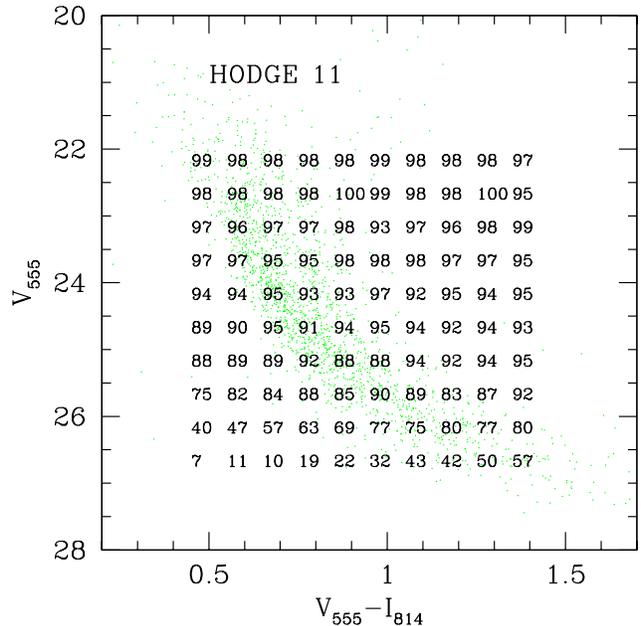}}
\caption[]{Observed colour-magnitude diagram of the field near Hodge 11 (dots).
 The number grid shows the $\mathrm{C(V-I,V)\times 100}$\% completeness values 
for the same field, calculated according to the method described in \S 2.3.} 
\label{h11_comp}
\end{figure}

Figure \ref{h11_comp} shows an example of this bi-variate completeness function
$\mathrm{C(V-I,V)}$. The completeness values are shown in percentage
values (or $\mathrm{C(V-I,V)\times   100}$\%). They are superposed
on the CMD itself. It is possible to observe that, at faint magnitude levels, 
$\mathrm{C(V-I,V)}$ decreases with V magnitude, as expected. 
For V \simlt 26, however, the data seem to be largely complete. 
Moreover, $\mathrm{C(V-I,V)}$ is larger
in the redder part of the low main sequence (for V \simgt 26). The latter
effect may be easily understood as part of the selection process: faint stars
in V, but bright in I (i.e., redder stars) have a larger 
probability to be selected than faint stars in both bands. This effect 
yields larger $\mathrm{C(V-I,V)}$ for redder V-I colours.

The bi-variate $\mathrm{C(V-I,V)}$ function is used as a probability that
a model star, generated according to the algorithm described in \S 4.1,
is actually included in the sample. The details are left for that
section. It is clear that the completeness function values suffer from
some uncertainty. Therefore, our CMD analysis shown in \S 5 was restricted 
to CMD regions where completeness was close to unity (typically 
$\mathrm{V} \simeq 24.5-25.0$). Notice that the lower parts of the
CMD are not crucial to the SFH reconstruction process, since the SFH is
more strongly influenced by main-sequence turn-offs and the positions
of evolved stars.

\section{Combining different samples}

In this work we have done the sample selection and 
photometry for 6 LMC fields. In Castro et al. (2001), 
7 independent (or almost) LMC fields were studied. The data from 
both studies have 
a common origin, since they were observed with the same instrument and
as part of the same project. All these fields are parallel HST/WFPC2 
observations of a rich LMC cluster and therefore
lie about 7.3 arcmin from this cluster. Therefore, our 6 fields can be
paired with 6 of the fields studied by Castro et al. (2001).
We thus decided to combine the stellar samples in each pair of fields
in order to enhance the number of stars used in the CMD analysis.

\begin{figure}[!h] 
\resizebox{\hsize}{!}{\includegraphics*[1.6cm,6cm][11cm,25cm]{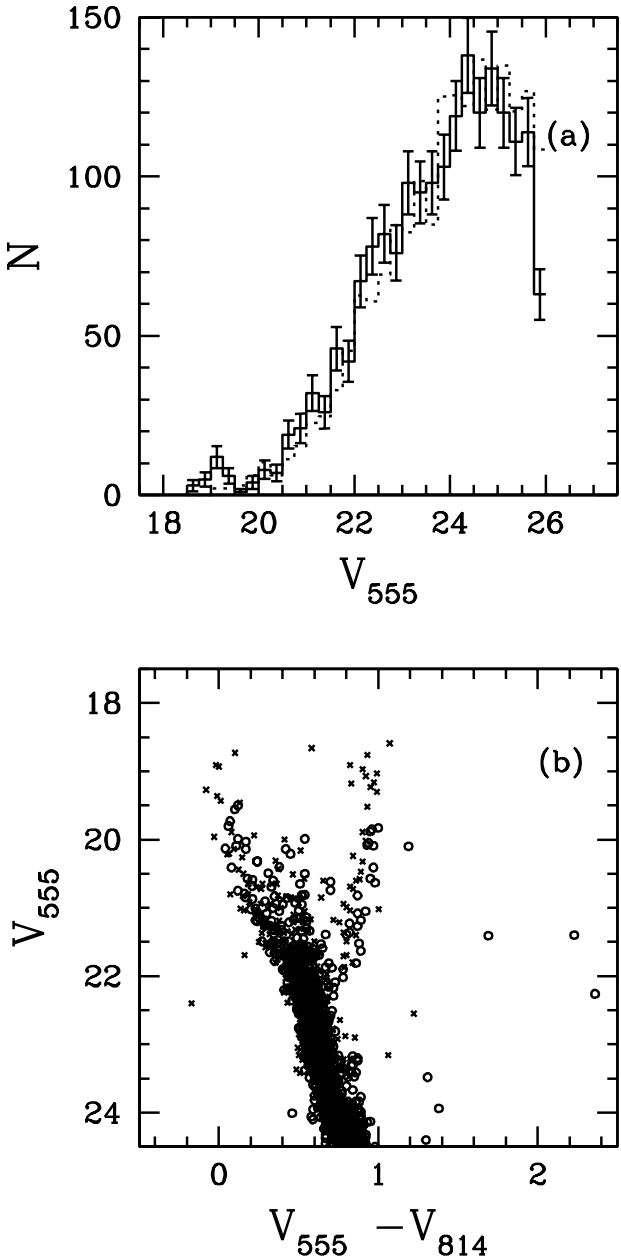}}
\caption[]{{\bf (a)} Luminosity functions of the field near NGC1831. 
Our sample is shown as a dotted line and that of Castro et al. (2001) 
 as a solid line with Poissonic bars. {\bf (b)} Observed colour-magnitude 
diagrams of our 
NGC1831 field (open circles) and that of Castro et al. (2001) (crosses).} 
\label{n1831_lf_cmd}
\end{figure}

In Figure \ref{n1831_lf_cmd}, we compare the stellar luminosity function (LF)
(upper panel) and the CMD (lower panel) of the two fields close to the rich
LMC cluster NGC1831. The LF from our sample (dotted line in panel 3a) has 
been corrected for completeness, as described in \S 2.3. The LFs are 
very similar, the
differences in star counts being in general consistent with random 
fluctuations ($\sqrt{N}$ bars are shown for the Castro et al. data). 
At the brightest magnitude bins ($V_{555}$ \simlt $19.5$), 
the excess of stars from Castro et al. reflects their brighter saturation 
limit, caused by their shorter exposure time. At intermediate
magnitudes, there may also exist a slight systematic excess of stars in the 
field studied by in Castro et al. (2001), suggesting that field-to-field
variations in number counts are not entirely random. Even though the
Castro et al. LF has not been corrected for sample incompleteness, this
effect manifests itself only fainter than $V_{555} \sim 25$. 
The CMDs (panel 3b) from both
samples match one another quite naturally, delineating the same loci.
The RGBs are slightly displaced from each other (by $\Delta (V_{555} - 
I_{814}) \sim 0.05$, in the sense of our data (open circles) being redder).
This behavior is not uncommon in the other fields as well and probably result
from PSF variations with time, which may in turn reflect optical
variations, such as changes in telescope focus. 

In order to place both samples in a common photometric system and thus 
avoid artificial spread in the combined sample CMD, we apply colour
offsets to Castro et al. data. This was done by first defining a fiducial line
for each separate sample, formed by the mean colour values at 0.25 
magnitude bins along the MS. We then changed the $I_{814}$ 
magnitude values of Castro et al. (2001) so that the two fiducial lines would
match each other. For the RGBs, we found a single colour difference and
changed the $\mathrm{I_{814}}$ magnitudes from Castro et al. to bring them together.
Typical corrections in colour were of $\sim 0.05$. Notice that this
approach, besides being simple, yields a homogeneous photometric dataset, 
while preserving as much as possible of the original ($\mathrm{V_{555}}$,
$\mathrm{V_{555} - I_{814}}$) CMD.

The final combined sample had to necessarily respect the differences 
in exposure time, sample selection and completeness corrections.
Exposures times of Castro et al. 2001 data were smaller than ours. 
Consequently, their CMDs have more information on the bright part of
the CMD, while our data have tend to better sample the faint part.
Also, they applied no completeness correction to their sample. 
Therefore, we defined the following combination scheme:
\begin{enumerate}

\item For the bright CMD region (V $< 19.5$) $\rightarrow$ only the
Castro et al. stars were used;

\item For the faint CMD (V $> 23.5$) $\rightarrow$ only our stars were 
included;

\item For the intermediate CMD ($19.5<$ V $< 23.5$) $\rightarrow$ both samples
 were included;

\end{enumerate}

The upper magnitude limit for our sample was meant to conservatively avoid 
any effect caused by saturation in our photometry.
Likewise, the lower limit for Castro et al. stars is a conservative
estimate of their completeness magnitude limits. The combination scheme
adopted, of course, significantly alters the number counts along the observed
CMD, since the effective solid angle covered by the sample is twice the size
in the intermediate CMD regions than in the upper or lower ones. 
In Sect. 4 we describe how to incorporate this sampling effect in a
model CMD. This is done simultaneously to the incorporation of completeness 
effects.

In two fields, the ones close to NGC1805 and NGC1831, some positional overlap
exists between the fields studied here and those by Castro et al. (2001). 
Duplication of stars was prevented by using only our data in these
overlapping regions. Positional matches between the two samples 
were found using the astrometry from the task METRIC available in the 
IRAF.STSDAS package. As the astrometric solutions for the overlapping fields 
were not unique, small positional offsets ($\simeq 0.5 \arcsec$) were
applied to the Castro et al. positions in order to make them comparable to 
ours.

Table \ref{tab2} shows the results of our sample combination: column 2
lists the number of stars in our sample alone, whereas column 3
lists the number of stars in the composite sample. Both numbers correspond
to stars brighter than the adopted completeness limit $\mathrm{V} \simeq 24.5-
25.0$. The final stellar sample is markably
greater than our original one, which translates into better
statistics and more reliable CMD modelling.

\begin{table}
\caption[]{The result of combining the two photometric samples as discussed in
\S 3. The columns show, respectively, field name, original number of stars 
in our sample and number of stars in the combined one.}
\label{tab2}
\small
\renewcommand{\tabcolsep}{0.6mm}
\begin{center}
\begin{tabular}{l c r c r}
\hline
Field &$~~~~~~$ & \#stars &$~~~~~~$ & \#stars  \cr
$~~~~$ & $~~~~~~$ & our sample &$~~~~~~$ & composed sample \\ \hline
NGC1805 &$~~~~~~$ & 2576&$~~~~~~$ & 3857 \cr
NGC1818 &$~~~~~~$ & 3566&$~~~~~~$ & 5578 \cr
NGC1831 &$~~~~~~$ & 1040&$~~~~~~$ & 1623 \cr
NGC1868 &$~~~~~~$ &  806&$~~~~~~$ & 1395 \cr
NGC2209 &$~~~~~~$ & 1019&$~~~~~~$ & 1483 \cr
Hodge 11& $~~~~~~$ & 1420 &$~~~~~~$ & 2143 \\ \hline
\end{tabular}
\end{center}
\hspace{3cm}
\end{table} 

\begin{figure}[!h] 
\resizebox{\hsize}{!}{\includegraphics*[1.cm,5cm][15cm,25cm]{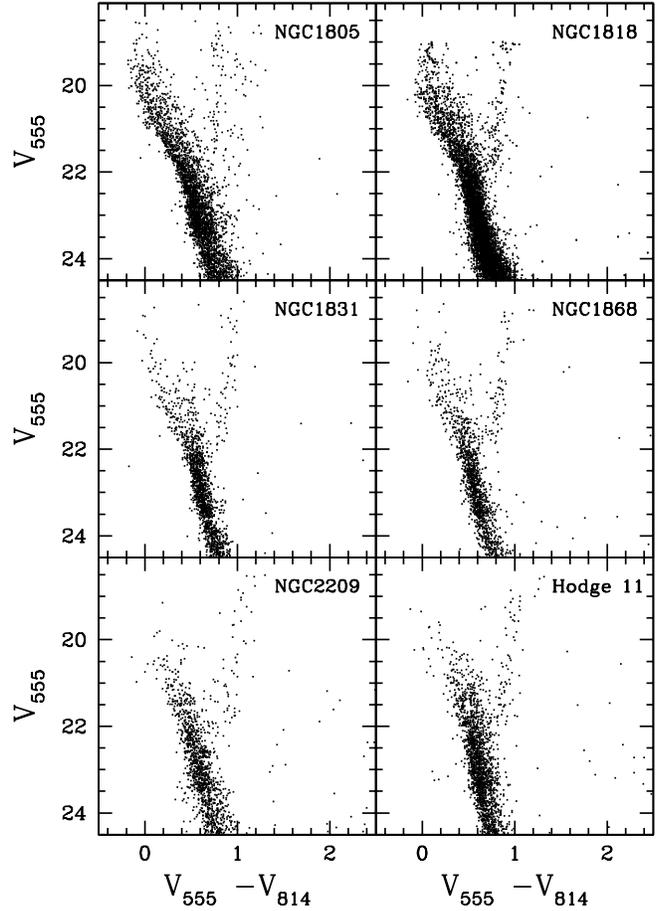}}
\caption[]{CMDs of the final combined samples for each of the fields, which are identified  by the rich cluster close to which they lie.} 
\label{CMDs_all}
\end{figure}

Figure \ref{CMDs_all} shows the CMDs of the final combined sample for
all 6 fields, which are again identified by the rich cluster
close to which they lie. These may be directly compared to the CMDs shown
by Castro et al. (2001) (figures 3 and 4). Apart from variations in 
magnitude limits, our final CMDs display the same features and have similar
MS and RGB positions and widths to those of the previous work. In particular,
the RGB in the NGC1805 region remains much broader than the other RGBs,
something that was interpreted as due to differential reddening.
Some candidate turn-offs of intermediate populations identified 
by Castro et al. (2001) still remain, while others have proven to be
artefacts caused by limited star counts.

\section{CMD modelling and statistical tools}

\subsection{Synthetic CMD}

CMDs have precious information concerning the star formation history
of a given population. From observed CMDs one can identify MS turn-offs of
various ages, therefore constraining the number of star forming bursts and
their ages, place limits on the youngest stars from MS termination
and also constrain the metal enrichment, based on isochrone fitting to
turn-offs and the RGB. Star counts in selected CMD areas are also useful
in discriminating star formation scenarios. A viable technique,
resulting from ever improving stellar evolution models and data quality,
is based on building theoretical CMDs according to a full model SFH.
By comparing observed CMDs to synthetic model ones,
we can then obtain constraints on the SFH that make use of all the 
information available. The process of generating artificial 
stars and building the synthetic CMD used in this work is as follows:
\begin{enumerate}

\item We choose a model SFH, by assigning relative star formation rates 
(SFRs) as a function of time, spanning the range of ages to be covered,
typically 8.00 \simlt $log(\tau)$ \simlt 10.25.

\item We also adopt an age-metallicity relation, $Z(\tau)$. In the
present analysis we chose:

\begin{itemize}
\item $Z(\tau)=0.008$ for $\tau$ \simlt 950 Myrs;
\item $Z(\tau)=0.006$ for 1 Gyr \simlt $\tau$ \simlt 2.1 Gyr;
\item $Z(\tau)=0.004$ for 2.2 Gyr \simlt $\tau $ \simlt 7 Gyr and
\item $Z(\tau)=0.002$ for 7.6 Gyr \simlt $\tau $ \simlt 12.6 Gyr and
\item $Z(\tau)=0.001$ for  $\tau$ \simgt 12.6 Gyr
\end{itemize}

This choice of $Z(\tau)$ is representative of the available data on LMC 
clusters (Olszewski et al. 1996).

\item We use an isochrone set that covers our age and metallicity intervals.
In this paper we have used the isochrones from Girardi et al. (2000); Given
a SFH ($SFR(t)$), each isochrone contributes with $N_{theo,i}$ stars,
given by: 

\begin{equation}
N_{theo,i}(\tau_{1},\tau_{2})=C \int_{\tau_{1}}^{\tau_{2}} SFR(\tau) \int_{m_{min}(\tau)}^{m_{max}(\tau)} \varphi(m)dm d\tau 
\label{ntheo}
\end{equation}
where the age limits $\tau_1$ and $\tau_2$ are chosen from the intervals in the
isochrone grid (in the case of Girardi et al. 2000, successive isochrones are 
placed $\Delta log \tau = 0.05$ apart). $\varphi(m)$ is the chosen IMF and
the limits in the mass integral are those that map onto the magnitude
limits of the observed CMD, using the mass-luminosity relation 
embedded in the isochrones themselves. For the IMF we used that of Kroupa
(2002), there being currently little variation in IMF shape inferred
for different populations (Kroupa 2002). The mass limits will depend not only
on the Padova stellar models, but also on the adopted distance modulus, since
our apparent magnitude CMD limits will map onto different absolute magnitude
limits as this latter parameter is varied. The intrinsic
distance modulus at each
field position was allowed to vary around a reference value taken 
from a model for the LMC disk inclined by 45$^o$ relative to the
line of sight, with the line of nodes aligned in the north-south direction
and $\mathrm{(m-M)_0 = 18.5}$ for the LMC centre
(Westerlund 1990, see also Castro et al. 2001 and 
\S 5 for more details).
$C$ is a constant of normalization, chosen so as to match the chosen total
number of model stars (given by $\sum_i N_{theo,i}(\tau_1,\tau_2)$).

\item Having found the number $N_{theo,i}$ of stars to be
drawn from the $i^{th}$ isochrone, we randomly draw stellar masses 
according to the IMF of Kroupa 
(2002) and compute the corresponding I and V magnitudes given by the 
isochrone and the chosen intrinsic $\mathrm{(m-M)_0}$. The masses are 
drawn as follows: given
a random number $r$, between 0 and 1, we solve the equation below for $m$:

\begin{equation}
\int_{m_{min}(\tau)}^{m}\varphi(m)dm =r\,
\int_{m_{min}(\tau)}^{m_{max}(\tau)}\varphi(m)dm d\tau
\end{equation}

\item We allow for unresolved binarism by repeating step (4), 
assuming that all binaries are coeval. We then combine the I and V 
fluxes of the two components and compute the corresponding system 
magnitude and colour. As the binary fraction,
$f_{bin}$, is largely unconstrained for LMC field stars, we 
explored different values of $f_{bin}$ and tested the sensitivity 
of our models to different choices of this parameter (see \S 5). 

\item We then apply the reddening vector $\mathrm{(A_{V},E(B-V))}$ to
the system magnitudes and colours, defining its theoretical CMD position. 
For this purpose we use $\mathrm{\frac{A_{V}}{E(B-V)}=3.1}$ and the 
photometric transformation to the vegamag WFPC2 system according to 
Holtzman et al. (1995a); the $\mathrm{E(B-V)}$ values primarily used were 
taken from Castro et al. (2001), although we also tried
alternative values of this parameter (see \S 5). 

\item We spread our star positions in the CMD based on a 
Gaussian distribution of errors with $\mathrm{\sigma_{V}}$ and 
$\mathrm{\sigma_{I}}$ as measured from PHOT/IRAF (see \S 2.2).

\item We verify whether the star (or binary system) ends up
inside the magnitude limits of observed CMD and discard it if it does not.

\item With the {\it observed } CMD position at hand, we then incorporate
the sampling effects. A probability $p$ of actually getting into the CMD is
assigned to each star and compared to a randomly chosen number $r$. If
$r \leq p$ the star is included, otherwise it is discarded. The
probability is $p = 0.5~\mathrm{C(V-I,V)}$ in the CMD regions where only
one of the photometric samples (either ours or from Castro et al.) was used,
and $p=\mathrm{C(V-I,V)}$ in the CMD regions where both samples contributed
to the observed CMD.

\end{enumerate}

\begin{figure}[!h] 
\resizebox{\hsize}{!}{\includegraphics*[1.5cm,6cm][20cm,25cm]{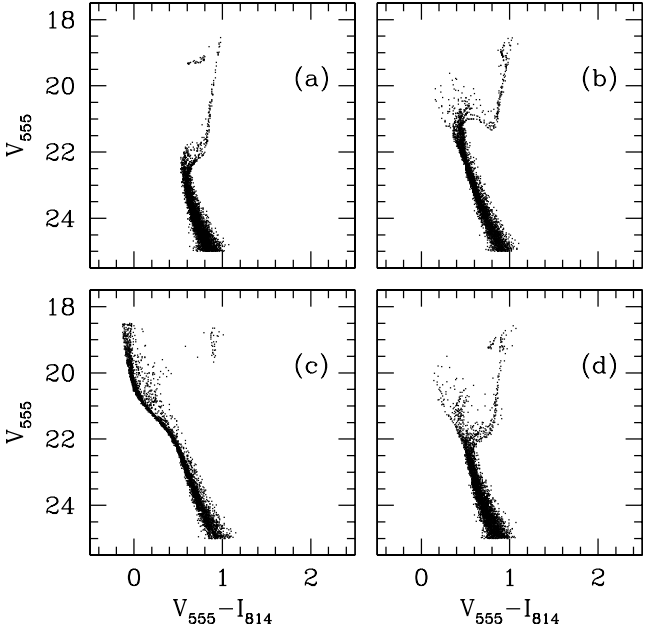}}
\caption[]{Some synthetic CMDs. {\bf (a)} a pure old population; {\bf (b)} 
a pure intermediate-age population; {\bf (c)} a predominantly young
population and {\bf (d)} a mixture of an intermediate-age and an old 
population.}
\label{sint_cmds}
\end{figure}

In Figure \ref{sint_cmds} we show some examples of synthetic
CMDs: {\bf (a)} a pure old population ($\tau$ \simgt 9 Gyr); {\bf (b)} a pure
intermediate-age population ($\tau \sim$  2 -- 3 Gyr); {\bf (c)} a 
predominantly young population ($\tau$ \simlt 1 Gyr) and {\bf (d)} a 
mixture of an intermediate-age and an old population ($\tau \sim$ 1 -- 3 Gyr 
and $\tau$ \simgt 6 Gyr). In figure \ref{sint_cmds}{\bf (a)} we see only
old stars and, as a consequence, the CMD shows a clear single turn-off whose
magnitude $\mathrm{V_{555}}$ is fainter and redder than those seen
in the other panels. Moreover panel {\bf (a)} shows a horizontal
branch, common to an old population CMD. A red clump is visible in panels 
{\bf (b)} and {\bf (d)}, as a consequence of intermediate-age stars. 
Panel {\bf (c)} is the only one whose main sequence stars extends
towards the brightest part of CMD ($\mathrm{V_{555}}$ \simlt 19), with
very few evolved stars.

\subsection{Statistical tools}

For CMD modelling we used a direct comparison of the distribution of stars in
the observed CMD with model CMDs resulting from a large number of
possible SFHs. The comparison method used was the Partial Models Method (PMM), 
developed by Gallart, Aparicio
and collaborators and described in detail by Gallart et al. (1999). Other
methods that reconstruct a SFH from direct CMD comparisons do exist 
(Hernandez, Valls-Gabaud \& Gilmore 1999, Dolphin
2002), some of which have been applied to LMC fields. Our choice of PMM was
based on the simplicity of implementing it.

The approach in the PMM is to create several synthetic CMDs from a 
constant $SFR(\tau)$, 
each CMD covering an age interval; these CMDs are called partial models. CMDs
from any function $SFR(\tau)$ can then be defined as a linear combination of 
the partial models. If we divide the CMD into a given number of
regions or boxes, the number of model stars expected to be found
in the $j^{th}$ CMD region is given by:

\begin{equation}
N_{s,j}=\sum _{i=1}^{m}a_{i}N_{i,j} 
\label{pmm_eq1}
\end{equation}

where $m$ is the number of partial models used, $N_{i,j} $ is the number of 
stars from partial model $i$ located in the CMD region mentioned. 
The set $a_{i}$ coefficients can be varied arbitrarily, 
accounting for any shape for the $SFR(t)$, with the only
constraint that 

\begin{equation}
\sum_{i=1}^{m}a_{i} = 1 
\label{pmm_eq2}
\end{equation}

The coefficients $a_i$ are relative SFR values, i.e., $a/<a>=SFR/<SFR>$. 
The partial models are created from a constant SFH and
by fixing the normalization constant $C$ in eq. \ref{ntheo} for all
age intervals $[\tau_1,\tau_2]$.
To obtain the SFH that best describes the data CMD,
we use essentially the $\chi_{\nu}^2$ statistical tool defined by 
Gallart et al. (1999), but with the
modification proposed by Mighell (1999) and Dolphin (2002), 
which is shown below:

\begin{equation}
\chi_{\nu} ^{2}=\frac{1}{\nu }\sum _{j=1}^{r}\frac{[l(\sum_{i=1}^{m} a_{i}N_{i,j})-N_{o,j}+1]^{2}}{N_{o,j}+1} 
\label{pmm_eq3}
\end{equation}
and 

\begin{equation}
\nu =r-1 
\label{pmm_eq4}
\end{equation}

In eq. \ref{pmm_eq3} $N_{o,j} $ is the number of observed stars in the 
$j^{th}$ CMD 
region , 
$r$ is the number of CMD regions and \( l \) is the normalization factor,
which is the ratio of the total numbers of observed to model stars.
The $\chi_{\nu}^2$ parameter allows one to discriminate, among the
different SFHs (represented by different sets of $a_i$), those
that best reproduce the observed data. 
We have to keep in mind the non-uniqueness of the best solutions. One needs
a criterion to select satisfactory models. We defined the acceptable range 
of $\chi_{\nu}^2$ values as being

\begin{equation}
(\chi_{\nu}^2)_{min}< \chi_{\nu}^2< (\chi_{\nu}^2)_{min} +2\sigma_{S}
\label{eq_cond}
\end{equation}
where $\sigma_{S}$ represents the standard deviation from 
$(\chi_{\nu}^2)_{min}$ of 100 realizations comparing the best model CMD (the 
solution that yields $(\chi_{\nu}^2)_{min}$) to the observed CMD. We 
considered the average $a_i$ values among these solutions as representing
the best SFH, using the scatter around this average as an uncertainty estimate.

\begin{figure}[!h] 
\centering 
\resizebox{\hsize}{!}{\includegraphics{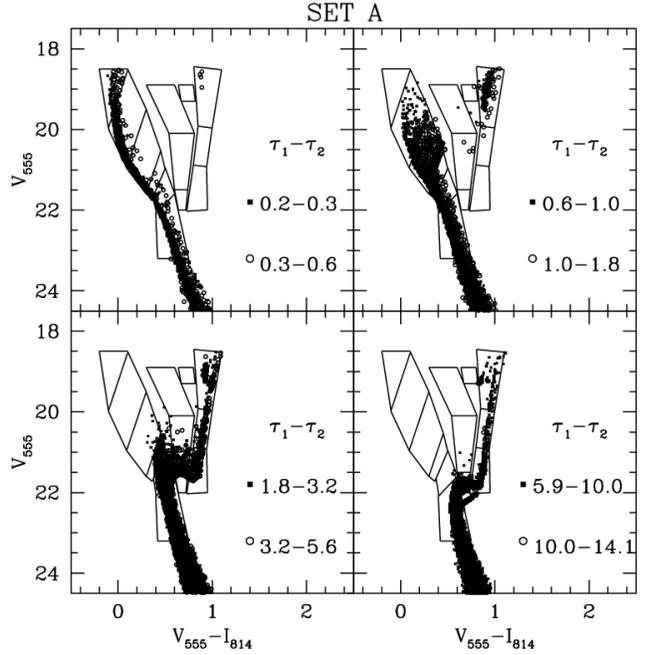}}
\caption[]{Set A - artificial CMD composed by 8 partial models. Age 
intervals are showed in Gyrs. Each panel shows two partial models. The regions used in the modelling process are 
indicated.}
\label{a_pmm}
\end{figure}

The choice of the number of CMD regions in the data $\times$ model comparison
was driven by the need to sample different CMD positions while keeping a 
significant number of stars in each region. We defined $r=13$ CMD regions, 
which efficiently cover
the MS, the subgiant branch (SGB) and the RGB, and which are large enough to
contain at least a few tens of stars in a typical observed CMD. These regions 
will therefore 
be able to detect the contributions from different populations, i. e., stars 
with different ages and chemical 
compositions. Figure \ref{a_pmm} shows the chosen CMD regions, along with
a synthetic CMD made up of $m=8$ partial models (2 partial models in 
each panel), whose age ranges are 
indicated. Most partial models are of younger ages. This choice is due to 
the fact that CMDs features are more sensitive to younger turn-offs 
(1 -- 6 Gyrs) than to older ones.  
We also tried other sets of partial models, with $m=7$ and $m=5$.
In appendix A we show the results of several control 
experiments which we used to test the method and its sensitivity to $m$.
As a result of these tests, we chose set A (with $m=8$) as the most capable 
of recovering an unbiased SFH.

\section{Results}
The CMD modelling method that we adopted followed a regular
parameter grid in the SFH: we considered all SFHs possible, by varying 
the $a_i$ ($i=1,8$) coefficients from 0 to 1, in steps of 0.1, and 
respecting the constraint given by eq. (\ref{pmm_eq2}). 
As for the other model parameters, it was impossible to fully explore parameter
space. Therefore, we kept the IMF fixed at all times to that 
from Kroupa (2002) and tested 4 values of E(B-V), 3 values of 
$\mathrm{f_{bin}=0.25,0.50,0.75}$ and 3 values of 
$\mathrm{(m-M)_0}$ (the one listed in Castro et al. 2001 and the other two 
by varying it by $\pm$ 0.1). In order to account for LMC depth effects,
individual model stars were assigned $\mathrm{(m-M)_0}$ values Gaussianly 
scattered by $\sigma_{(m-M)_0} = 0.02$ around the model distance. This 
represents
$\simeq \pm  500 pc$ distance spread relative to LMC disk mid-point. 
We considered as our {\it best} solution for 
$\mathrm{E(B-V)}$, $f_{bin}$, $\mathrm{(m-M)_0}$ in each field the one that 
resulted in an absolute $\chi_{\nu}^2$ minimum in the entire parameter grid. 
Our best SFH solution was then the average $a_i$ values among those which 
satisfied the criterion given by eq. \ref{eq_cond}. We tested the sensitivity 
of the SFH solution (as well as its $(\chi_{\nu}^2)_{min}$ value) to 
$\mathrm{E(B-V)}$, distance modulus and $f_{bin}$. The results are shown 
in \S 5.4. 

Figures \ref{sfh1} to \ref{sfh3} show the SFHs solutions 
recovered for the different LMC fields, all of them expressed in terms 
of $SFH/<SFH>$. As mentioned before, we took the average $SFR/<SFR>$ values 
within the criterion 
given by eq. (\ref{eq_cond}) and the 1$\sigma$ deviations from the average as 
uncertainty bars. A common feature in all recovered SFHs is the existence 
of a non-negligible old ($\tau > 10 Gyrs$) population, as had been previously 
pointed out by Holtzman et al. (1997), Geha et al. (1998), among others. Most 
SFHs deviate significantly from uniformity, displaying periods of 
enhanced star formation ($SFR/<SFR>$ \simgt 2) and of quiescence 
(where $SFR/<SFR>$ \ltsima 0.5). 

\subsection{NGC1805 and NGC1818}

\begin{figure}[!t] 
\resizebox{\hsize}{!}{\includegraphics*[1cm,8.5cm][20.5cm,22cm]{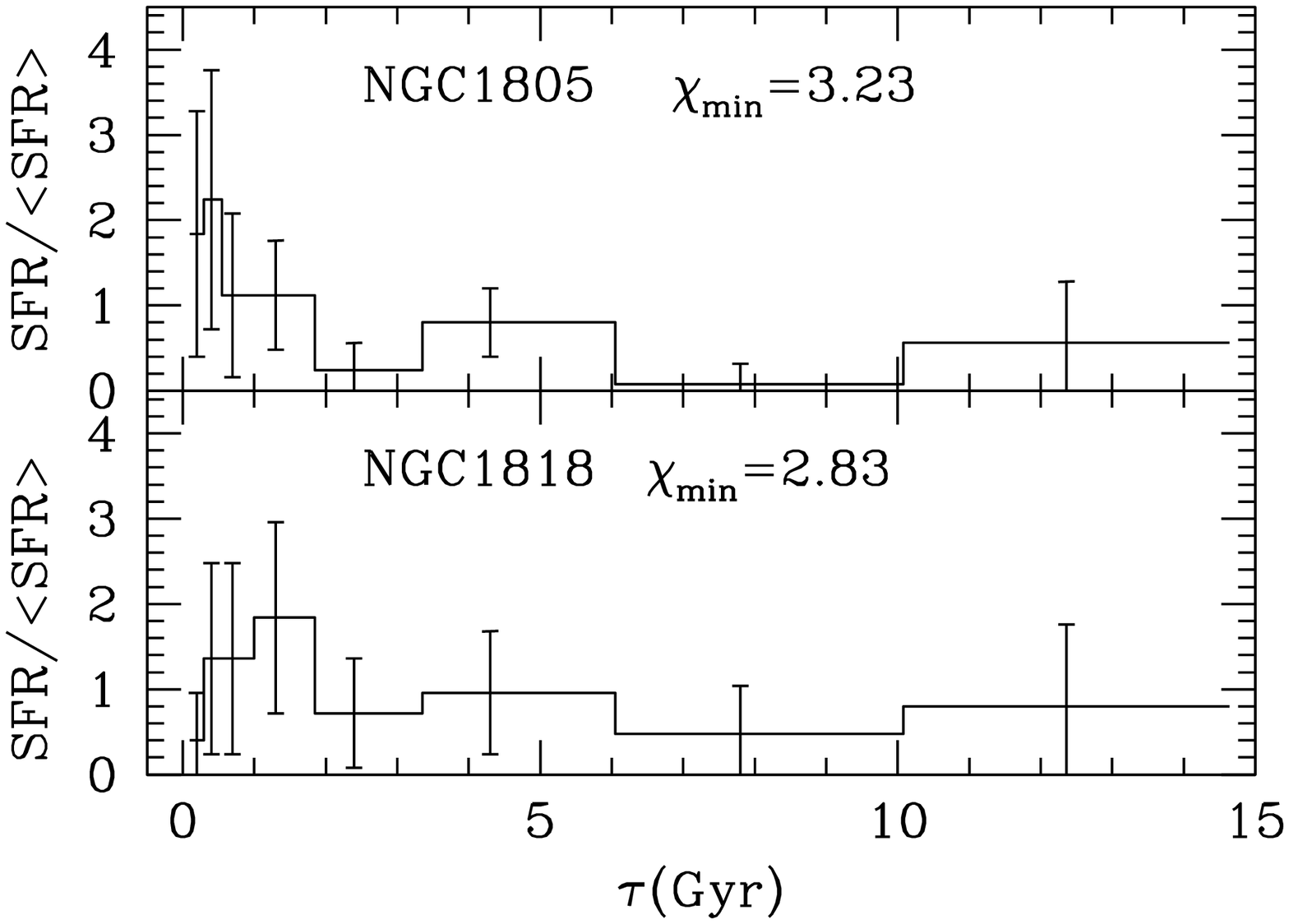}}
\caption[]{The star formation histories (SFHs) recovered from NGC1805 
(top panel) and NGC1818 (bottom panel) fields, using the
E(B-V) values that yield the minimum $\chi_{\nu}^2$ (E(B-V) = 0.03 and 
E(B-V) = 0.05 for NGC1805 and NGC1818, respectively). The lines show 
the average of
all values of $\frac{a_{i}}{\bar{a}}$ which satisfy the criterion given
by eq. (\ref{eq_cond}). The error bars show the 1$\sigma$ scatter from the average.} 
\label{sfh1}
\end{figure}

In figure \ref{sfh1}, top panel, we show our {\it best} SFH solution for 
the field close to NGC1805, obtained with $\mathrm{E(B-V)=0.03}$, 
$f_{bin} = 0.75$ and $\mathrm{(m-M)_0 = 18.59}$. 
The lower panel shows the result for the field close to NGC1818.
The absolute minimum $(\chi_{\nu}^2)_{min}$ for NGC1818 was found for 
E(B-V)=0.05, $f_{bin} = 0.5$ and $\mathrm{(m-M)_0 = 18.58}$.
Both SFHs depicted in Figure \ref{sfh1} are quite distinct from the
{\bf uniform case}. Relatively recent episodes of star formation
have occurred in the past $\sim 2$ Gyrs. A broad and significant gap in
star formation is present in NGC1805 at $6 - 10$ Gyrs
ago.  In NGC1818, our results point to a star formation closer to uniform 
before $2$ Gyrs ago; the apparent decrease in SFR in the interval 
$6 - 10$ Gyrs is still present, but is marginally significant.

\subsection{NGC1831 and NGC1868}

\begin{figure}[t] 
\resizebox{\hsize}{!}{\includegraphics*[1cm,8.5cm][20.5cm,22cm]{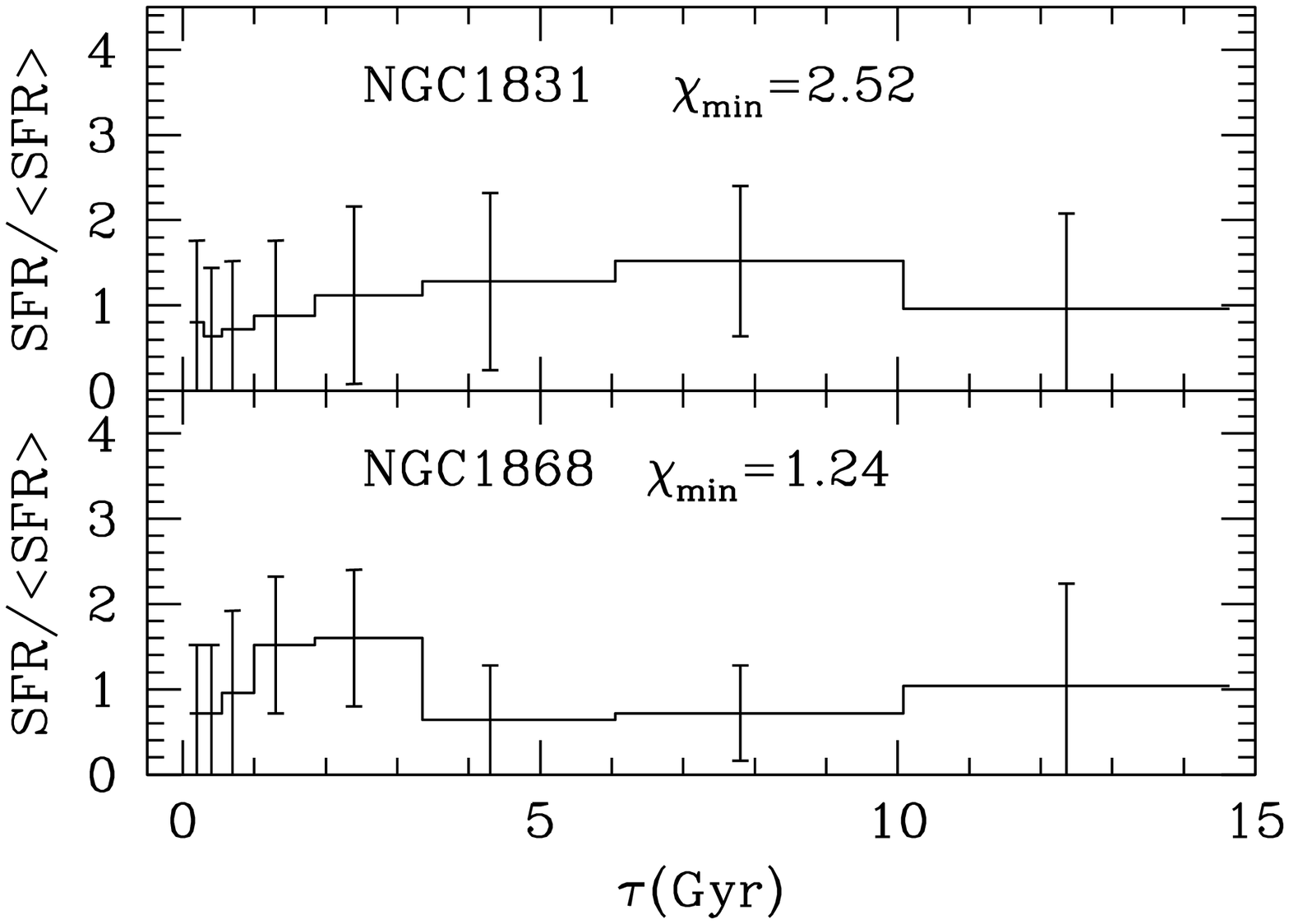}}
\caption[]{The star formation histories (SFHs) recovered from NGC1831 
(top panel) and NGC1868 (bottom panel) fields, using the
E(B-V) values that yield the minimum $\chi_{\nu}^2$ (E(B-V) = 0.03 and 
E(B-V) = 0.00 for NGC1831 and NGC1868, respectively). The lines show 
the average of all values of $\frac{a_{i}}{\bar{a}}$ which satisfy 
the criterion given
by eq. (\ref{eq_cond}). The error bars show the 1 $\sigma$ scatter from this
average.}  

\label{sfh2}
\end{figure}

The best SFH solutions for the fields close to NGC1831 and
NGC1868 are drawn in Figure \ref{sfh2}. The recovered SFHs for both fields are
more consistent with uniformity than the previous two. The quiescent
period between $6 - 10$ Gyrs ago is marginally seen in the NGC 1868
solution and is totally absent in the NGC 1831 field SFH.
In the NGC1868 field an increase in star formation in the past
$\tau \simeq 1$ -- $3$ Gyrs is also apparent. The best values for 
distance modulus 
and binary fraction for this field were $\mathrm{(m-M)_0}$ = 18.45 and 
$f_{bin} = 0.50$, respectively. For NGC1831, we found $\mathrm{(m-M)_0}$ = 
18.48 and $f_{bin} = 0.75$. The $\mathrm{E(B-V)}$ values corresponding to 
$(\chi_{\nu}^2)_{min}$: E(B-V)= 0.03 for NGC1831 and E(B-V)=0.00
for NGC1868.
 
\subsection{NGC2209 and Hodge 11}

\begin{figure} [t] 
\resizebox{\hsize}{!}{\includegraphics*[1cm,8.5cm][20.5cm,22cm]{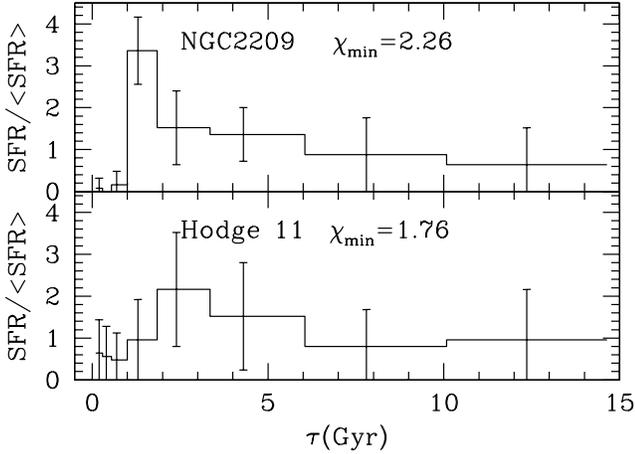}}
\caption[]{The star formation histories (SFHs) recovered from NGC2209 
(top panel) and Hodge 11 (bottom panel) fields, using the
E(B-V) values that yield the minimum $\chi_{\nu}^2$ (E(B-V) = 0.11 and 
E(B-V) = 0.10 for NGC2209 and Hodge 11, respectively). The lines show 
the average of all values of $\frac{a_{i}}{\bar{a}}$ which satisfy the 
criterion given by eq. (\ref{eq_cond}). The error bars show the 1 $\sigma$ 
scatter from this average.}  
\label{sfh3}
\end{figure}

In figure \ref{sfh3} one can find the results for the NGC2209 and Hodge
11 fields. For the first we found E(B-V)=0.11, $\mathrm{(m-M)_0}$ = 
18.39 and $f_{bin} = 0.5$ based on the absolute $(\chi_{\nu}^2)_{min}$ 
values, whereas for Hodge 11, our minimization procedure yielded 
E(B-V) = 0.10, $\mathrm{(m-M)_0}$ = 18.34 and $f_{bin} = 0.25$. 
The best SFH solutions for these two fields are similar. They both point
to an increase in star formation within the interval $1 - 6$ Gyrs ago,  
especially in the case of NGC 2209, where a significant peak is seen at
$\tau \simeq 1$ Gyr. Star formation seems to have subdued significantly since
then, according to these SFH solutions.

\begin{figure} [h] 
\resizebox{\hsize}{!}{\includegraphics{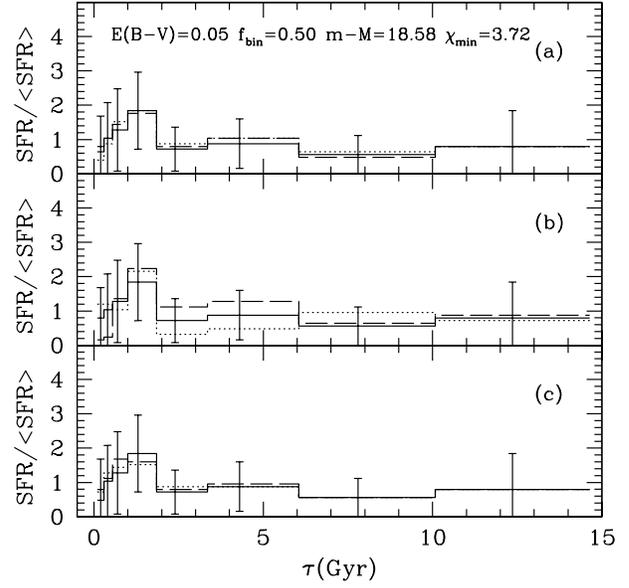}}
\caption[]{Different star formation histories (SFHs) recovered from the field 
near NGC1818, resulting from adoption of different $\mathrm{E(B-V)}$ values (panel 
{\bf (a)}), $\mathrm{(m-M)_0}$ (panel {\bf (b)}), $f_{bin}$ (panel {\bf (c)}). 
{\bf (a)} The different lines correspond to different choices of 
$\mathrm{E(B-V)}$: solid - $\mathrm{E(B-V)} = 0.05$; dotted - 
$\mathrm{E(B-V)} = 0.03$; dashed - $\mathrm{E(B-V)} = 0.04$. {\bf (b)} The 
different lines show solutions with different values adopted for 
$\mathrm{(m-M)_0}$: solid - $\mathrm{(m-M)_0} = 18.58$; 
dotted - $\mathrm{(m-M)_0} = 18.68$; dashed - $\mathrm{(m-M)_0} = 18.48$. 
{\bf (c)} The different lines 
show solutions with different values adopted for $f_{bin}$: solid - 
$f_{bin} = 0.5$; dotted - $f_{bin} = 0.75$; dashed - $f_{bin} = 0.25$. In 
all cases, they correspond to the average of all values of 
$\frac{a_{i}}{\bar{a}}$ which satisfy the criterion given by 
eq. (\ref{eq_cond}). The error bars show the 1 $\sigma$ scatter from 
this average.}  
\label{sfh_ebv}
\end{figure}

\begin{figure}  
\resizebox{\hsize}{!}{\includegraphics{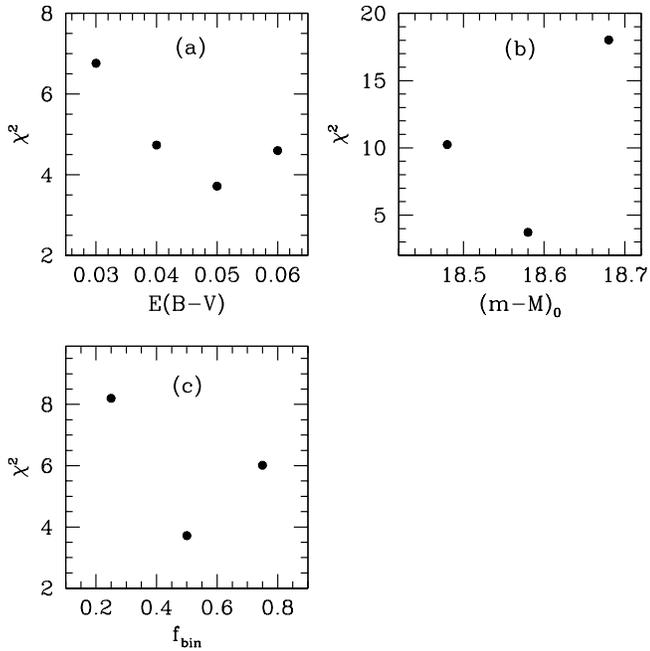}}
\caption[]{$(\chi_{\nu}^{2})_{min}$ values recovered from the field 
near NGC1818, resulting from adoption of different $\mathrm{E(B-V)}$ (panel 
{\bf (a)}), $\mathrm{(m-M)_0}$ (panel {\bf (b)}), $f_{bin}$ (panel {\bf (c)}). 
}  
\label{chi_par}
\end{figure}

\begin{table}[t]
\caption[]{The best $\mathrm{E(B-V)}$ are shown in columns $\mathbf{a-d}$ 
while  $\mathrm{(m-M)_0}$ values are listed in columns $\mathbf{e,f}$. 
{\bf (a, e)} - our results; {\bf (b, f)} - Castro et al. (2001) results; 
{\bf (c)} values quoted by Santiago et al. (2001) for LMC clusters. 
{\bf (d)} - Burstein \& Heiles (1982) values. }
\label{ebv}
\small
\renewcommand{\tabcolsep}{2mm}
\begin{center}
\begin{tabular}{l c c c c c c}
\hline
Field  & \multicolumn{4}{c}{$\mathrm{E(B-V)}$}  & \multicolumn{2}{c}{$\mathrm{(m-M)}$}  \\ \cr
$~~~$  & $a$ & $b$ & $c$  &  $d$ & $e$  & $f$ \\ \hline
NGC1805  & 0.03 & 0.04 & 0.04 & 0.03 & 18.69 & 18.59 \cr
NGC1818  & 0.05 & 0.04 & 0.03& 0.03 & 18.58 & 18.58 \cr
NGC1831  & 0.03 & 0.0 & 0.015 & 0.01  & 18.48 & 18.58 \cr
NGC1868  & 0.0 & 0.02 & 0.0 & 0.01 & 18.45 & 18.55 \cr
NGC2209  & 0.11 & 0.07 & 0.11 & 0.11 & 18.39 & 18.39 \cr
Hodge 11  & 0.10 & 0.0 & -- & 0.09 &18.34 & 18.34 \\ \hline
\end{tabular}
\end{center}
\hspace{9cm}
\end{table}

\subsection{Testing the best solution parameters}

Our method of CMD modelling yields non-unique solutions for the SFH in a
given field; different sets of parameter values often result in acceptable
solutions.  Figure~\ref{sfh_ebv} shows the SFH solutions in the field 
close to NGC1818 corresponding to different choices of $\mathrm{E(B-V)}$ 
(panel {\bf a}), $\mathrm{(m-M)_0}$ (panel {\bf b}) and $f_{bin}$ 
(panel {\bf c}). 
The best values of E(B-V), $f_{bin}$, $\mathrm{(m-M)_0}$ and the resulting 
absolute $(\chi_{\nu}^2)_{min}$ are quoted on the upper panel. In each
panel, we show the sensitivity of the SFH solution to varying one 
grid parameter at a time. The figure shows that the SFHs vary modestly
when model parameter values close to the best solution are considered.
Changes in $\mathrm{(m-M)_0}$ tend to be more important when one wants to 
find a meaningful solution for the SFH. Figure \ref{chi_par} shows 
$(\chi_{\nu}^2)_{min}$ obtained for NGC1818 with the same parameters
whose SFH solutions are presented in figure \ref{sfh_ebv}. One
can see the behavior of $(\chi_{\nu}^2)_{min}$ with respect to 
$\mathrm{E(B-V)}$ (panel {\bf a}), $\mathrm{(m-M)_0}$ (panel {\bf b}) and 
$\mathrm{f_{bin}}$ (panel {\bf c}). It is remarkable that a global 
$(\chi_{\nu}^2)_{min}$ was found in this case. The same behaviour
is also observed in the other fields.

As described in the previous sections, 
our choice of {\it best solution} was the one which yielded the absolute 
minimum in $(\chi_{\nu}^2)$ with respect to $\mathrm{E(B-V)}$, 
$\mathrm{(m-M)_0}$ and $f_{bin}$. We also tested our CMD modelling approach 
by comparing our best $\mathrm{E(B-V)}$ and $\mathrm{(m-M)_0}$ to other 
values of the same parameter, found independently by other methods. 
Castro et al. (2001) value of distance modulus was based on a model for the 
LMC disk, as described in \S 4.1. Their extinction values were based
on isochrone fits to the CMDs of their sample of field LMC stars. 
Santiago et al. (2001) also applied isochrone fits, but to the observed CMDs of
the nearby {\it rich LMC clusters}. Since the clusters are 7.3 arcmin from 
the fields studied here, we assume that any variation in extinction across 
this distance is negligible. Notice that this assumption may well be wrong 
in the case of NGC1805, where the broad RGB, seen both in the field and 
cluster CMDs, may indicate differential redenning.  
$\mathrm{(m-M)_0} = 18.50$ was assumed in these isochrone fits for all 
clusters. Finally, we also use the $\mathrm{E(B-V)}$ values from the
Burstein \& Heiles (1982) maps, which are based on galaxy counts, for
comparison.

Table \ref{ebv}  shows our best $\mathrm{E(B-V)}$ (column $\mathbf{a}$) 
and $\mathrm{(m-M)_0}$ (column $\mathbf{e}$) values provided by the partial 
models method. In the other columns we show Castro et al. (2001) values for
$\mathrm{E(B-V)}$ (column $\mathbf{b}$) and $\mathbf{(m-M)_0}$ (column 
$\mathbf{f}$). The alternative $\mathrm{E(B-V)}$ value by Santiago et al. 
(2001) is listed in column $\mathbf{c}$. Both works used Padova isochrones 
from Girardi et al. (2000) in their fits. 
Burstein \& Heiles (1982) $\mathrm{E(B-V)}$ values appear 
in column $\mathbf{d}$. 

There is an excellent agreement among the different determinations
of extinction and distance modulus for these fields, especially if we 
consider the different methods and assumptions that are involved.
The only exception is the very low extinction value adopted by Castro et al.
(2001) for the field close to Hodge 11, which, according to those authors, 
should be considered cautiously as a possible result of a wrong distance 
modulus or mean metallicity adopted for the data.

Finally, we also carried out the same CMD modelling analysis with a cruder
version of the chemical-enrichment law presented here in order to test the
sensitivity of SFHs to the assumed chemical enrichment. The SFHs show
only mild changes but the main characteristics remain, including the
existence of field-to-field variations in star formation.

\section{Summary and conclusions}

In this work we have analyzed 6 deep CMDs of field stars
in the LMC using HST/WFPC2 data in the $V_{555}$ and $I_{814}$ filters. 
Our fields are located from 1.5$^{\circ}$ to 6$\degr$ from the LMC centre 
and within $\simeq \, 7'$ of rich LMC clusters. From 800 to 3500 stars were 
measured in both V and I bands in each field. We combined these data with those
obtained by Castro et al. 2001, carefully placing both datasets in a 
homogeneous photometric system, which resulted in a much larger photometric
sample. 

We have studied the behavior of photometric uncertainties and 
of sample incompleteness in the CMDs. We developed, 
tested and applied a method to correct for photometric incompleteness as a 
function not only of magnitude V, but also of (V-I) color. 

Based on the final observed CMDs, and with the aid of synthetic model CMDs
and objective statistical methods, we have constrained the LMC star formation 
history. Our synthetic CMDs are capable of incorporating observational 
effects like extinction, distance modulus, photometric errors and unresolved 
binarity. For CMD modelling we used the Partial Models Method (Gallart et al. 
1999). We have tested the method through control experiments using 3 different 
sets of partial models. We also examined the sensitivity of our results to 
adopted distance modulus, binary fraction and extinction. 

Although the uncertainty bars are considerable, the recovered SFHs vary 
from one field to another. The two fields closer
to the LMC bar, NGC1805 and NGC1818, show less uniform SFHs, with
stronger events of star formation. Star formation peaks are seen in the
past $\sim 2 Gyrs$. A relatively quiescent 
phase from $\sim 6$ to $\sim 10$ Gyrs ago is also observed, with 
significant star formation earlier on. Periods of enhanced
star formation at intermediate  ages ($\sim 2 - 6$ Gyrs) are 
observed in the fields on the eastern side of the LMC, NGC2209 e 
Hodge 11. The recent star forming activity ($\tau < 1$ Gyr) 
seen in NGC1805 and NGC1818 is no longer observed. Finally, a more uniform 
star formation history results from CMD modelling of the fields in the NW part 
of the LMC disk, NGC1831 and NGC1868, especially in this first. 
The only important variation from 
a uniform SFR in NGC1868 is possibly an enhanced star formation period from $\tau \simeq 1$ to 
$\tau \simeq 3$ Gyrs. In all fields, one
can clearly observe the significant contribution of older populations
($\tau$ \gtsima $6$ Gyrs) to their CMDs.

In brief, we can partially relate the recovered SFHs to fields position: 
fields near the LMC bar (NGC1805 and NGC1818) present more recent star 
formation ($\tau$ \simlt $1-2 Gyrs$), something
that is actually confirmed by a larger number of upper MS stars on their 
observed CMDs. NGC2209 and Hodge 11, located to the east of the LMC
center, show little recent star formation, but display an excess of
intermediate age ($\tau \sim 2 - 6$ Gyrs) stars. 
Finally, NGC1831 and NGC1868 are a further
step closer to a uniform SFH, although some increase in star formation in the 
last $\simeq 3$ Gyrs is also visible in the NGC 1868 field.

It is important to recognize that the inferred SFHs do not describe the
star formation rates that took place {\it in situ}, for the stars currently
found in our HST/WFPC2 fields have certainly dispersed away from their 
various formation sites. 
Therefore, the periods of enhanced formation we find should reflect 
star formation rates over much broader regions of the LMC, perhaps the
entire galaxy in the case of the older populations. The same
applies to the intervals of lower than average activity. The observed
differences from one field to the other show that the mixing
of stellar populations within the LMC field is not 100\% efficient; on
the contrary, there exist coeval stars, resulting from the same burst, 
which move coherently in the galaxy's potential, regardless of their ages or
original location. A dynamical model, taking into account typical velocity
dispersions and orbital motions, is then required in order to
reconstruct the star formation sites, burst time-scales and locations, from
our resulting SFHs.

This work gives continuity to a historical process of recognition of
the importance of old and intermediate-age populations in the LMC. Until the
beginning of the 80s, the LMC was seen as being
formed of relatively young stars (Butcher 1977,  Stryker 1984). In the
beginning of the 90s, Bertelli et al. (1992) and Vallenari et al. (1996)
suggested a star formation burst at $\tau \sim $ 4 Gyr and
found evidence for some spatial variation in the star formation.
The HST observations revealed a
considerable portion of intermediate-age and old stars. The recent
reconstructed SFHs lead us to conclude that stars older than
4 Gyr are at least as numerous as younger ones (Holtzman et al. 1997, 
Geha et al. 1998). Our results clearly corroborate this notion.

Another important result is that the gap in star formation in the LMC is
not seen in the SFHs derived here; evidence for a decrease in the SFR 
in the interval $6 - 10$ Gyrs is seen only in the fields closer to the 
LMC bar. It is therefore hard to reconcile the SFH of fields stars with the
observed age distribution of LMC clusters, which reveals a longer and older
age gap.
  
It is important to remember that we have always used the same chemical
enrichment model which is consistent with metallicities and ages of LMC
clusters (Olszewski et al. 1996). However, there is evidence,
including from the present work, that the field stars SFH differs
from the SFH of the LMC clusters (Holtzman et al. 1999). Thus, it
would be extremely important to obtain direct and strong constraints 
on the chemical enrichment of
LMC field stars (Cole et al. 2000, Smecker-Hane et al. 2002). For this
challenge one needs the new generation telescopes, such as 
GEMINI and SOAR.

\begin{acknowledgements}
We acknowledge the Coordena\c c\~ao de Aperfei\c coamento de Pessoal 
do N\'\i vel
Superior (CAPES) and the Conselho Nacional de Desenvolvimento Cient\'\i fico
e Tecnol\'ogico (CNPq) for partially supporting this work, the latter 
through the PRONEX/FINEP/CNPq program 76.97.1003.00. We acknowledge 
David Valls-Gabaud for useful discussions and providing us with many of 
the codes used in our analysis. Finally, we thank the referee, Dr. Gallart, for
her useful sugestions and comments. 

\end{acknowledgements}

\bibliographystyle{aa}

\appendix
\label{app}

\section{Control Experiments}

\begin{figure*}
\centering 
\includegraphics[width=17cm]{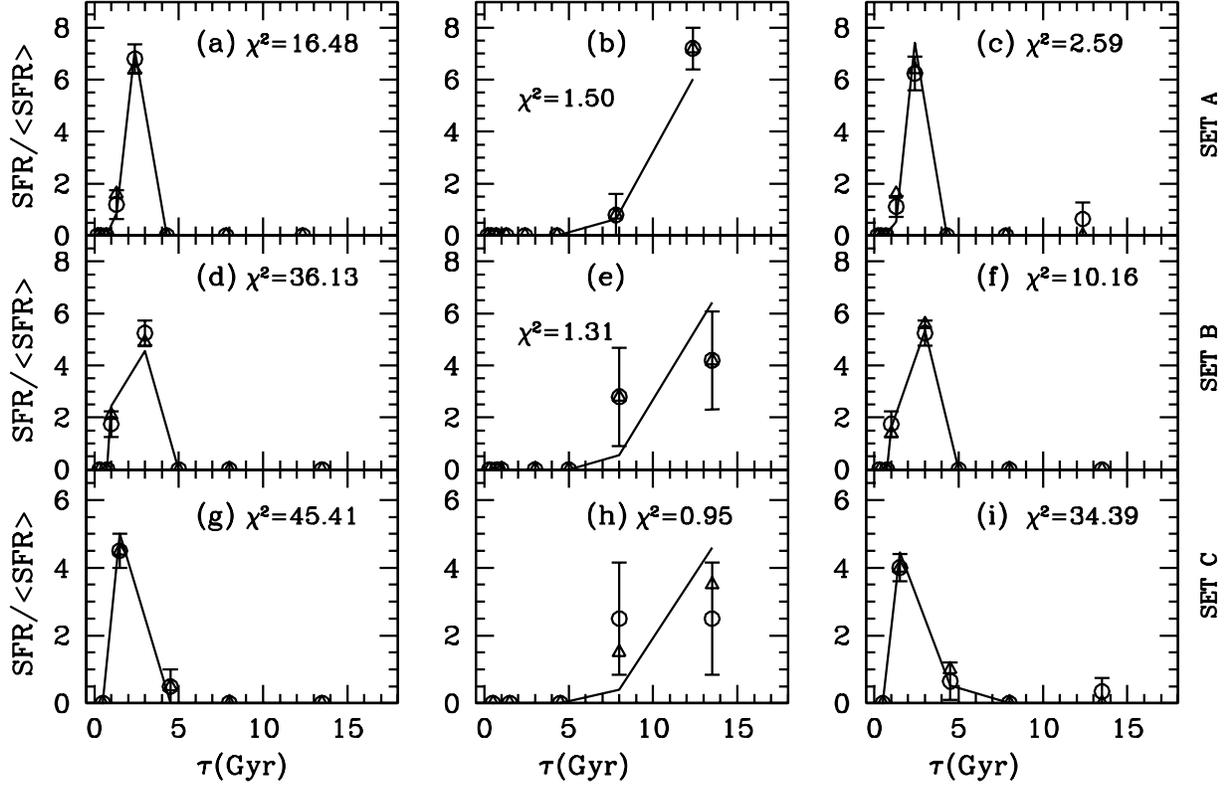}
\caption[]{Each column shows SFH schemes given in $\mathrm{SFR/<SFR>}$ (solid lines) and corresponding recovered SFH (open circles). Each line shows the results for each set of partial models: set A (panels {\bf (a)}, {\bf (b)} and {\bf (c)}); set B (panels {\bf (d)}, {\bf (e)} and {\bf (f)}) and set C (panels {\bf (g)}, {\bf (h)} and {\bf (i)}). The set of $a_{i}$ values obtained on best solution (i. e. $(\chi_{\nu}^2)_{min}$). Error bars show acceptable $a_{i}$ values with dispersion within $1\sigma$. } 
\label{ce1}
\end{figure*}

\begin{figure*}
\centering 
\includegraphics[width=17cm]{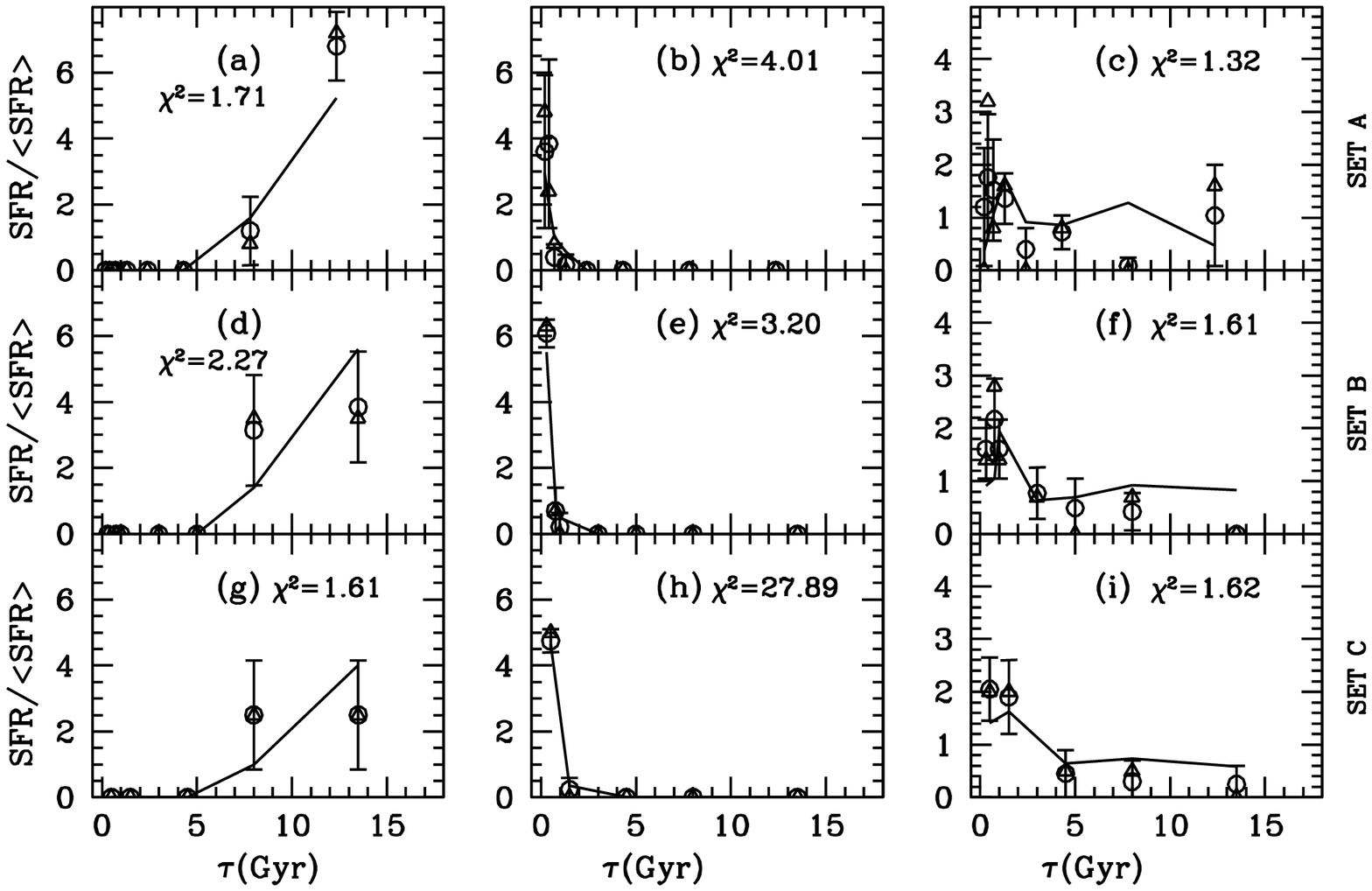}
\caption[]{Same as shown in figure \ref{ce1}, but for the other 3 SFH schemas.}
\label{ce2}
\end{figure*}

\begin{figure*}
\centering 
\includegraphics[width=17cm]{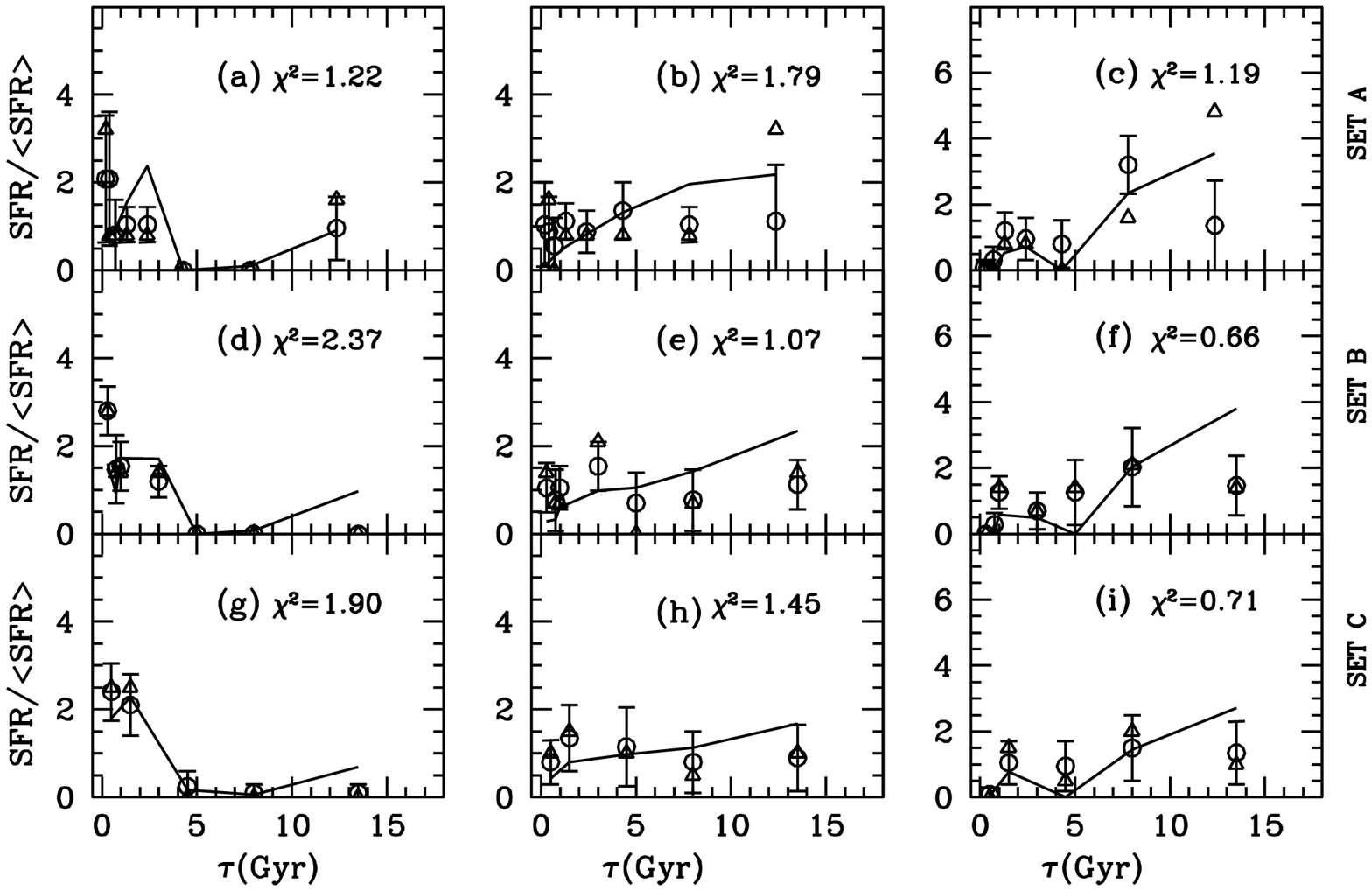}
\caption[]{Same as shown in figure \ref{ce1}, but for the other 3 SFH schemas.}
\label{ce3}
\end{figure*}

In order to implement the PMM, we need to define the number of partial models 
to use, $m$, and to check the efficiency of the method. This is done by 
generating an
artificial CMD based on some known input SFH, as described in Sect. 4.1, 
and submitting this CMD to 
the modelling process described in Sect. 4.2. In other words, we compare
this ``artificial data'' CMD with other artificial CMDs created in the same
way but following alternative schemes of SFH. This comparison makes use of the
PMM, with different choices of $m$.
The results of such control experiments can be viewed in figures 
\ref{ce1}--\ref{ce3}. In these figures, the panels along each horizontal 
line refer to the results of a given set of 
partial models. We tested three sets: A ($m=8$), B ($m=7$) and C ($m=5$).
The vertical columns of the panels correspond to a given input SFH.
The input SFH is given as a solid line in each panel. It varies slightly
from one set to the other because of the variation in the 
age intervals covered by each set.
The {\it best solution}, as described in \S 5, and which uses the 
quality criterion given in \S 4.2, is shown as open circles. 
The open triangles correspond to the sets $\{a_{i} \} $ that yield 
$(\chi_{\nu}^2)_{min}$, i.e., the minimum $(\chi_{\nu}^2)$ value 
among all $\{a_{i} \} $ sets. This $(\chi_{\nu}^2)_{min}$ value is 
given in each panel. 
The figures show a total of 9 input SFHs, 3 per figure.

In most cases, the recovered SFHs (either the best or minimum solution) 
are very similar to the input ones, which shows that the method is in fact
capable of reconstructing a SFH from CMD data. However, some discrepancies 
significantly beyond the error bars are seen in a few points, especially in 
panels A.2.c, A.2.f, A.3.a, A.3.d, A.3.f. In most these cases,
only one point along the SFH is not well recovered, as is the case of the underestimate
in star formation rate (SFR) at $\tau \simeq 8$ Gyrs in panel A.2.c.
This latter panel actually shows a situation (for $\tau \simeq 0.5$ Gyr) 
where the best solution is a good description of the input model, whereas the
minimal solution is not. This is also seen in panels A.3.b and A.3.e. We thus conclude that the adoption of an average solution 
over some range in $(\chi_{\nu}^2)$ close to the minimum, as in our best solution, rather
than the minimum itself, results in more stable results.
Another conclusion is that SFHs dominated by a single burst tend to be more
easily reproduced, while more complex or nearly constant star formation input 
scenarios often result in recovered SFHs with misplaced peaks 
(A.3.a, A.3.c and A.3.f) or slightly distorted SFH shapes (as in A.2.c, 
A.3.b, A.3.e, A.3.h).
Based on the general results from these control experiments, and considering
the relative frequency of discrepancies between input and recovered SFRs, we
decided to use set A of partial models, with $m = 8$; this set shows about
the same number of discrepant points as set B ($m=7$) and set C ($m=5$), but
samples better the SFH.

\end{document}